\documentclass[11pt,a4paper]{article}
\usepackage{amsmath,amssymb}
\usepackage{epsfig,graphicx}
\usepackage[english]{babel}
\usepackage{verbatim}
\usepackage{cite}

\begin{document}

\title{\bf Chiral color symmetry \\ 
and  $G'$-boson mass limit from Tevatron~data~on~$t \bar t$-production}
\author{M.~V.~Martynov$^a$\footnote{{\bf e-mail}: martmix@mail.ru},
A.~D.~Smirnov$^{a}$\footnote{{\bf e-mail}: asmirnov@uniyar.ac.ru}
\\
$^a$ {\small Division of Theoretical Physics, Department of Physics,}\\
{\small Yaroslavl State University, Sovietskaya 14,}\\
{\small 150000 Yaroslavl, Russia.}}

\date{}
\maketitle

\begin{abstract}
\noindent
A gauge model with chiral color symmetry of quarks is considered and possible effects
of the color $G'$-boson octet predicted by this symmetry are investigated. 
The contributions of the $G'$-boson to the cross section $\sigma_{t\bar{t}}$ 
and to the forward-backward asymmetry $A_{\rm FB}^{p \bar p}$ of $t\bar{t}$ production 
at the Tevatron are calculated and  analysed in dependence on two free parameters of the model,
the mixing angle $\theta_G$ and $G'$ mass $m_{G'}$.
The $G'$-boson contributions to $\sigma_{t\bar{t}}$ and $A_{\rm FB}^{p \bar p}$ are shown
to be consistent with the Tevatron data on $\sigma_{t\bar{t}}$ and $A_{\rm FB}^{p \bar p}$,
the allowed region in the $m_{G'} - \theta_G$ plane is discussed
and around  $m_{G'}=1.2 \, TeV, \; \theta_G=14^\circ $
the region of~$1 \sigma$ consistency is found.

\vspace{5mm}


\end{abstract}




The search for a new physics beyond the Standard Model (SM) is now one
of the aims of the high energy physics. 
The simplest extentions of the SM (such as two Higgs models, models based on supersymmetry,
left-right symmetry, four color quark-lepton symmetry or models implying the four fermion generation, etc.)
predicting the new physics effects at one or a few TeV energies are most interesting now in anticipation
of the new results from the LHC which will allow the investigations of new physics effects
at the TeV energy scale with very large statistics \cite{Butterworth:2007bi}.

One of the simplest extentions of the SM can be based on the idea of the originally chiral character
of $SU_c(3)$ color symmstry of quarks. i.e on the gauge group of the \underline{chiral color symmetry}
\begin{eqnarray}
\label{chiral_group}
\label{eq:Gcc}
G_c &=& SU_L(3)\!\times \! SU_R(3)  \to  SU_c(3), \\
    & & \hspace{5mm}       g_L \, , \hspace{9mm} g_R \hspace{3.5mm}   \to \hspace{4mm}  g_{st}, 
\nonumber 
\end{eqnarray}
which is assumed to be valid at high energies and is broken to usual QCD $SU_c(3)$ at low energy scale.
The immediate consequence of the chiral color symmstry of quarks is the prediction of 
\underline{new color-octet gauge particle}: the axigluon $G^A_\mu$ in the case 
of $g_L=g_R$~\cite{Pati:1975ze,Hall:1985wz,Frampton:1987ut,Frampton:1987dn} 
or the $G'$-boson in general case 
of~$g_L\neq g_R$~\cite{Cuypers:1990hb,Martynov:2009en,Mart_Sm_MPLA_2010}.      
\begin{eqnarray}
\hspace{-0mm}
 G_c \;\; \Rightarrow \;\; \left\{ 
\begin{array}{l} 
     \mbox{ axigluon} \; G^A \;\;  \mbox{for} \;\;  g_L=g_R,
\mbox{~\cite{Pati:1975ze,Hall:1985wz,Frampton:1987ut,Frampton:1987dn}}, \\
 G'-\mbox{boson} \;\; \mbox{for} \;\;  g_L\neq g_R, \;\;\;\; 
\mbox{~\cite{Cuypers:1990hb,Martynov:2009en,Mart_Sm_MPLA_2010}}, \\ 
\end{array}
\right.
\nonumber
\end{eqnarray}

The $G'$-boson is the octet-colored gauge particle with vector and axial vector coupling constants to quarks 
of order $g_{st}$ which are defined by gauge coupling constants $g_L, \,\,g_R$.     
Some features of the axigluon (including its phenomenology at the Tevatron) were investigated in 
ref.~\cite{Bagger:1988,frederix-2009,rodrigo-2008,antunano-2007} and 
the massive color octet with arbitrary vector and axial vector  coupling constants to quarks     
has been considered phenomenologicaly in ref.~\cite{ferrario-2008}.

Since it is the colored gauge particle with vector and axial vector coupling to quarks, 
the $G'$-boson should give rise the increase of the cross section as well as 
the appearance of a forward-backward asymmetry in $Q \bar Q$ production.

\underline {The current CDF data} on cross section $\sigma_{t\bar{t}}$ ~\cite{CDF9913} 
and forward-backward asymmetry $A_{\rm FB}^{p \bar p}$ ~\cite{CDFAFB2009} 
 of the $t\bar{t}$ production at the Tevatron are 
\begin{eqnarray}
\sigma_{t\bar{t}} & = & 7.5 \pm 0.31 (stat) \pm 0.34 (syst) \pm 0.15 (lumi) pb \, (= 7.5 \pm 0.48 \, pb) ,
\label{expcspptt09}
\\
A_{\rm FB}^{p \bar p} & = & 0.193 \pm 0.065~(\rm{stat}) \pm 0.024~(\rm{sys})\, (= 0.193 \pm 0.069). 
\label{AFBpptt09}
\end{eqnarray}

\underline{The SM predictions} for $\sigma_{t\bar{t}}$ and $A_{\rm FB}^{p \bar p}$ have been discussed in
refs.~\cite{Cacciari:2008zb,Kidonakis:2008mu,Moch:2008ai}
and~\cite{Kuhn:1998kw,Bowen:2005ap,antunano-2007,Almeida:2008ug}
respectively and we quote here the next SM predictions for $\sigma_{t\bar{t}}$~\cite{Cacciari:2008zb}
and $A_{\rm FB}^{p \bar p}$~\cite{antunano-2007}
\begin{eqnarray}
\sigma_{t\bar{t}}^{SM} & = & 7.35 {~}^{+0.38}_{-0.80}~\mathrm{(scale)}
{~}^{+0.49}_{-0.34} ~\mathrm{(PDFs)} \mathrm{[CTEQ6.5]}  \, \mathrm{pb}  \div
\label{sppttSM}
\\
\notag &\phantom{\div}& 7.93 {~}^{+0.34}_{-0.56}~\mathrm{(scale)}
{~}^{+0.24}_{-0.20} ~\mathrm{(PDFs)} \mathrm{[MRST2006nnlo]} \, \mathrm{pb} ,
\\
  A_{\rm FB}^{SM}(p \bar p \to t \bar t )  &=&  0.051(6).
\label{AFBppttSM}
\end{eqnarray}
The first and second values in~\eqref{sppttSM} were obtained in
NLO+NLL approximation with $m_t=171$~GeV and correspond to the different
choises of the parton distribution functions (CTEQ6.5 and MRST2006nnlo respectively).
As seen the experimental and theoretical values of $\sigma_{t\bar{t}}$~\eqref{expcspptt09},~\eqref{sppttSM}
are compatible within the experimental and theoretical errors
whereas the experimental value of $A_{\rm FB}^{p \bar p}$~\eqref{AFBpptt09} exceeds
the corresponding theoretical prediction~\eqref{AFBppttSM} by more than ~$2 \sigma$ .
This deviation is not so large nevertheless this circumstance is under active discussion 
now~\cite{Ferrario:2009bz, Zerwekh:2009vi,
Frampton:2009rk,Shu:2009xf,Ferrario:2009ee,Cao:2010zb, Dorsner:2009mq,Jung:2009jz,Djouadi:2009nb,
Cheung:2009ch,Arhrib:2009hu,Barger:2010mw,Cao:2009uz}.

The main goal of my talk is to clean up if the gauge chiral color symmetry~\eqref{eq:Gcc} is consistent 
with the CDF data~\eqref{expcspptt09}, ~\eqref{AFBpptt09} and what bounds on the mass of $G'$-boson are 
imposed by these data.  

In the case of the gauge chiral color symmetry~\eqref{eq:Gcc} 
the $3\times3$ matrices of the usual gluon fields $G_\mu$ and of the $G'$-boson fields $G'_\mu$ 
are constructed from the basic gauge fields $G^L_\mu$ and $G^R_\mu$ as   
\begin{align}
G_\mu&=  s_G \, G^L_\mu + c_G \, G^R_\mu,
\nonumber 
\\
G'_\mu&=  c_G \, G^L_\mu - s_G \, G^R_\mu,
\nonumber 
\end{align}
where 
\begin{eqnarray}
s_G =\sin\theta_G = \frac{g_R}{\sqrt{(g_L)^2+(g_R)^2}},  \,\,\,\,
c_G =\cos\theta_G = \frac{g_L}{\sqrt{(g_L)^2+(g_R)^2}},
\label{eq:sGcG}
\nonumber 
\end{eqnarray}
$\theta_G$ is $G^{L} - G^{R}$ mixing angle,  
$G_{\mu} = G^i_{\mu} t_i$, $G'_{\mu} = G'^i_{\mu} t_i$,   
$t_i$, $i=1,2,...,8$, are the generators of $SU_c(3)$ group. 

To reproduce the usual quark-gluon interaction of QCD the gauge coupling constants $g_L, \, g_R$
of the gauge group~$G_c$ must satisfy the relation
\begin{eqnarray}
\frac{g_L g_R}{\sqrt{(g_L)^2+(g_R)^2}} = g_{st}(M_{chc}). 
\label{eq:gLgRgst}
\nonumber 
\end{eqnarray}
where $M_{chc}$ is the mass scale of the chiral color symmetry breaking 
and $g_{st}(M_{chc})$ is the strong coupling constant taken at this mass scale.

The \underline {interaction of the $G'$-boson with quarks} in this case takes the form 
\begin{equation}
\mathcal{L}_{G'qq}=g_{st}(M_{chc}) \, \bar{q} \gamma^\mu (v + a \gamma_5) G'_\mu q , 
\label{eg:LG1qq}
\end{equation}
where $v$ and $a$ are the vector and axial-vector coupling constants for the which    
the gauge chiral color symmetry group $G_c$ gives  the expressions
\begin{equation}
v = \frac{c_G^2-s_G^2}{2 s_G c_G} = \cot(2\theta_G), \,\,\,\,
a = \frac{1}{2 s_G c_G} = 1 / \sin(2\theta_G) .
\label{eg:va}
\nonumber 
\end{equation}

As a result of the chiral color symmetry breaking the $G'$-boson picks up the mass  
\begin{equation}
m_{G'} = \frac{g_{st}(M_{chc})}{s_G c_G} \, \frac{\eta}{\sqrt{6}},
\label{eq:MG1}
\nonumber 
\end{equation}
where $\eta$ is the VEV of the $(3_L, \bar{3}_R)$ scalar field $\Phi_{\alpha \beta}$ of the group $G_c$, 
which breaks the chiral color symmetry,    
$\langle \Phi_{\alpha \beta} \rangle =  \delta_{\alpha \beta} \, \eta /(2 \sqrt{3}) $, \,
$\alpha, \beta = 1,2,3$ are the $SU_L(3)$ and $SU_R(3)$ indices.

So, the gauge chiral color symmetry model has two free parameters, the $G'$-boson mass $m_{G'}$ and   
the $G^{L} - G^{R}$ mixing angle $\theta_G$, $tg\,\theta_G=g_R/g_L$, 
which gives the possibility to study the phenomenology of the $G'$-boson in more detail 
in dependence on these two parameters.

The \underline {differential cross section} of the process 
$q\bar{q}  \stackrel{\,g,\,G'}{\rightarrow}  Q \bar{Q}$ 
in tree approximation with account of the $G'$-boson 
interaction~\eqref{eg:LG1qq} and of the gluon contributions 
has the form~\cite{Mart_Sm_MPLA_2010} 
\begin{eqnarray}
\nonumber
&&\frac{ d\sigma(q\bar{q} \stackrel{\,g,\,G'}{\rightarrow} Q \bar{Q}) }{d\cos \hat{\theta}} = 
 \frac{\pi \beta}{9\hat{s}}
\bigg \lbrace \alpha_s^2(\mu) \, f^{(+)} +
\frac{\alpha_s(\mu) \, \alpha_s(M_{chc}) \, 2 \hat{s} (\hat{s}-m_{G'}^2)}
{(\hat{s}-m_{G'}^2)^2+m_{G'}^2 \Gamma_{G'}^2}
\Big[ \, v^2 f^{(+)} + 2 a^2 \beta c \, \Big] +
\\ \label{diffsect}
&& + \frac{\alpha_s^2(M_{chc}) \, \hat{s}^2} {(\hat{s}-m_{G'}^2)^2+m_{G'}^2 \Gamma_{G'}^2}
\Big[ \left( v^2 + a^2 \right)
\big( v^2 f^{(+)}+   a^2 f^{(-)} \big)
+ 8 a^2v^2 \beta c \, \Big]
\bigg \rbrace,
\end{eqnarray}
where $f^{(\pm)}=(1+\beta^2 c^2\pm 4m_Q^2/\hat{s})$, $c = \cos \hat{\theta}$,
$\hat{\theta}$ is the  scattering angle of $Q$-quark in the parton center of mass frame,
$\hat{s}$ is the squared invariant mass of $Q \bar{Q}$ system,
$\beta = \sqrt{1-4m_Q^2/\hat{s}}$, $M_{chc}$ is  the mass scale of the chiral color symmetry breaking
and $\mu$ is a typical scale of the process.

The corresponding to~\eqref{diffsect} \underline {total cross section} 
takes the form~\cite{Mart_Sm_MPLA_2010}
\begin{eqnarray}
\nonumber
\sigma(q\bar{q} \stackrel{\,g,\,G'}{\rightarrow} Q \bar{Q}) &=& \frac{4\pi \beta}{27\hat{s}}
\bigg \lbrace
\alpha_s^2(\mu) \, (3-\beta^2) +
\frac{2\, \alpha_s(\mu)  \alpha_s(M_{chc})\, v^2   \hat{s}(\hat{s}-m_{G'}^2)(3-\beta^2)}
{(\hat{s}-m_{G'}^2)^2+\Gamma_{G'}^2 m_{G'}^2}+\\
&+&\frac{\alpha_s^2(M_{chc}) \, \hat{s}^2 \big[ \, v^4(3-\beta^2) + v^2 a^2 (3+\beta^2) +
2a^4\beta^2 \, \big]}
{(\hat{s}-m_{G'}^2)^2 + \Gamma_{G'}^2 m_{G'}^2}
\bigg \rbrace.
 \label{sect}
\end{eqnarray}

The enterring into~\eqref{diffsect},~\eqref{sect} 
\underline {hadronic width of the $G'$-boson} is known~\cite{ferrario-2008,Martynov:2009en} 
and can be written as
\begin{eqnarray}
\Gamma_{G'} =  \sum_{Q} \Gamma (G' \to Q\overline{Q})
\label{GammaG'}
\nonumber
\end{eqnarray}
where
\begin{eqnarray}
&&\Gamma (G' \to Q\overline{Q}) =
= \frac{\alpha_{s}(M_{chc}) \, m_{G'}}{6}
\Bigg[ \, v^2 \left(1+\frac{2m_Q^{2}}{m_{G'}^{2}}\right)
+ a^2 \left(1-\frac{4m_Q^{2}}{m_{G'}^{2}}\right) \Bigg] \sqrt{1-\frac{4m_Q^{2}}{m_{G'}^{2}}}
\label{GammaG'QQ}
\nonumber
\end{eqnarray}
is the width of $G'$-boson decay into $Q\overline{Q}$-pair.

At $M_{chc}=1.2 \, TeV$, for example,  we obtain the next estimations for the relative width of $G'$-boson
\begin{equation}
  \Gamma_{G'}/m_{G'}=0.08, \; 0.14, \; 0.33, \; 0.60, \; 1.37
\label{gammaG1}
\nonumber
\end{equation}
for  $ \theta_G=45^\circ, \; 30^\circ, \; 20^\circ, \; 15^\circ, \; 10^\circ $
respectively.

As concerns the process of $Q \bar{Q}$ production in gluon fusion $g g  \rightarrow Q \bar{Q}$  
the $G'$-boson does not contribute, in tree approximation, to this process. 

The \underline {differential and total partonic cross sections}
of the process of $Q \bar{Q}$ production in gluon fusion $g g  \rightarrow Q \bar{Q}$   
in tree approximation of the SM are well known and have the form
\begin{eqnarray}
&& \hspace{-10mm} \frac{d\sigma^{SM}_0(gg\rightarrow Q \bar{Q})}{d\cos \hat{\theta}} =
\alpha_s^2(\mu) \: \frac{\pi \beta}{6 \hat{s}}
\left(\frac{1}{1-\beta^2c^2}-\frac{9}{16}\right)
\left(1 + \beta^2 c^2 +2(1-\beta^2)-\frac{2 (1-\beta^2)^2}{1-\beta^2 c^2}\right),
\label{difcsggQQ}
\\[5mm]
&& \hspace{-10mm} \sigma^{SM}_0(gg\rightarrow Q \bar{Q}) = 
\frac{\pi  \alpha_s^2(\mu) }{48 \hat{s}}
\left[
\left(\beta ^4-18 \beta ^2+33\right) \log \left(\frac{1+\beta }{1-\beta }\right)+
\beta  \left( 31 \beta ^2-59 \right)
\right].
\label{totcsggQQ}
\end{eqnarray}

The $G'$-boson can generate, at tree-level, a \underline {forward-backward asymmetry} 
in $Q\bar{Q}$-pair production due to the \underline {forward-backward difference}
in the $q\bar{q}\to Q\bar{Q}$ partonic cross section~\cite{Mart_Sm_MPLA_2010}   
%
\begin{eqnarray}
&&\Delta_{FB}(q\bar{q}\to Q\bar{Q}) \hspace*{-1mm} =  \hspace*{-1mm} \sigma(q\bar{q}\rightarrow Q \bar{Q},
 \, \cos \theta > 0) \hspace*{-1mm} - \hspace*{-1mm}
\sigma(q\bar{q}\rightarrow Q \bar{Q}, \, \cos \theta < 0) \hspace*{-1mm} =
\nonumber
\\
&& = \frac{4\pi \beta^2 a^2}{9}
\Bigg( \frac{\alpha_s(\mu) \, \alpha_s(M_{chc}) \, (\hat{s}-m_{G'}^2) + 2 \alpha_s^2(M_{chc}) \, v^2 \hat{s}}
{(\hat{s}-m_{G'}^2)^2+m_{G'}^2 \Gamma_{G'}^2}
\Bigg), 
\label{eq:deltaFBqq}
\end{eqnarray}
which can give rise to the corresponding forward-backward asymmetry $A_{\rm FB}^{p \bar p}$
of $t\bar{t}$-pair production in $p \bar p$ collisions at the Tevatron.


\begin{figure}[htb]
\vspace*{0.5cm}
 \centerline{
\epsfxsize=0.55\textwidth
\epsffile{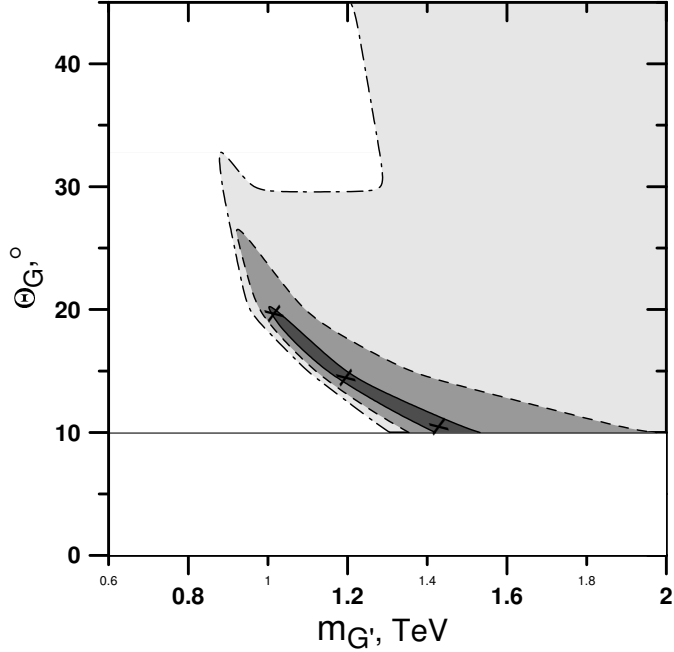}}
\vspace*{1mm}
\caption{The $m_{G'} - \theta_G$ regions consistent with CDF data
on cross section~$\sigma_{t\bar{t}}$  and forward-backward asymmetry~$A_{\rm FB}^{p \bar p}$
in $t \bar{t}$ production within $1 \sigma$ (dark region), $2 \sigma$ (grey region)
and $3 \sigma$ (light-grey region).}
\label{constrTevatron}
\end{figure}


We have calculated the \underline{cross section $\sigma(p \bar{p} \rightarrow t \bar{t})$} 
of $t\bar{t}$-pair production in $p\bar{p}$-collisions at the Tevatron energy
using the total parton cross section of quark-antiquark annihilation~\eqref{sect},
the total SM parton cross section~\eqref{totcsggQQ} of the gluon fusion $g g  \rightarrow Q \bar{Q}$
and the parton densities AL'03~\cite{alekhin} (NLO, fixed-flavor-number,  $Q^2=m_t^2$)
with the appropriate K-factor $K=1.24$~\cite{campbell-2007-70}.
Here and below we beleive $\mu^2=Q^2$, $M_{chc}=m_{G'}$.

With the same parton densities we have calculated the \underline{forward-backward
asymmetry $A_{\rm FB}^{p \bar p}$} in $t\bar{t}$-pair production at the Tevatron in the form
\begin{eqnarray}
    A_{\rm FB}^{p \bar p} & = & A_{\rm FB}^{G'} + A_{\rm FB}^{SM},
\label{AFBppttcalc}
\end{eqnarray}
where $A_{\rm FB}^{G'}$ is the corresponding $G'$ boson contribution which has been calculated
using the differential parton cross section~\eqref{diffsect}
(one can use also the expression~\eqref{eq:deltaFBqq}) and $A_{\rm FB}^{SM}$ is the SM prediction for
$A_{\rm FB}^{p \bar p}$ for which we have used the value~\eqref{AFBppttSM} of ref.~\cite{antunano-2007}.

We have analysed the cross section $\sigma(p \bar{p} \rightarrow t \bar{t})$ 
and the forward-backward asymmetry $A_{\rm FB}^{p \bar p}$
 in dependence on two free parameters of the model,
the mixing angle $\theta_G$ and $G'$ mass $m_{G'}$, in comparision with 
the Tevatron data~\eqref{expcspptt09},~\eqref{AFBpptt09}
on $\sigma_{t\bar{t}}$ and $A_{\rm FB}^{p \bar p}$. 
The result of this analysis is shown in $m_{G'} - \theta_G$ plane in Fig.1.

The Fig.1 shows the regions in the $m_{G'} - \theta_G$ plane which are simultaneusly consistent
with the data~\eqref{expcspptt09} and~\eqref{AFBpptt09} within $1 \sigma$ (dark region),
$2 \sigma$ (grey region) and $3 \sigma$ (light-grey region).
As seen from the Fig.1 for
\begin{equation}
m_{G'} \gtrsim 1.0 \, TeV
\label{eq:mG1limcsFAB}
\nonumber
\end{equation}
in the $m_{G'} - \theta_G$ plane there is the region which is
consistent with the CDF data~\eqref{expcspptt09},~\eqref{AFBpptt09} 
on $\sigma(p \bar{p} \rightarrow t \bar{t})$ and $A_{\rm FB}^{p \bar p}$.  

For example, for the masses
%
\begin{equation}
a) \; m_{G'} = 1.02 \, TeV, \; \; b) \; m_{G'} = 1.2 \, TeV, \; \; c) \; m_{G'} = 1.4 \, TeV
\label{eq:mG1values}
\end{equation}
with the appropriate values of $\theta_G$
($\theta_G=19^\circ, \, \theta_G=14^\circ, \, \theta_G=11^\circ$ respectively,
these points are marked in Fig.1 by crosses)
we obtain for $\sigma_{t\bar{t}}$, $A_{\rm FB}^{p \bar p}$ the values
\begin{eqnarray}
&& a)  \; \sigma_{t\bar{t}} = 7.98 \, pb, \; \; A_{\rm FB}^{p \bar p} = 0.158 \, (0.107),
\label{pointa}
\\
&& b)  \; \sigma_{t\bar{t}} = 7.61 \, pb, \; \; A_{\rm FB}^{p \bar p} = 0.154 \, (0.103),
\label{pointb}
\\
&& c)  \; \sigma_{t\bar{t}} = 7.57 \, pb, \; \; A_{\rm FB}^{p \bar p} = 0.141 \, (0.090),
\label{pointc}
\end{eqnarray}
which are consistent with the CDF data~\eqref{expcspptt09},~\eqref{AFBpptt09} 
on $\sigma(p \bar{p} \rightarrow t \bar{t})$ and $A_{\rm FB}^{p \bar p}$ within~$1 \sigma$.

In parentheses in~\eqref{pointa}-\eqref{pointc}   
 we show for comparision the $G'$-boson contributions in $A_{\rm FB}^{p \bar p}$
defined by~\eqref{eq:deltaFBqq}, without the SM contribution~\eqref{AFBppttSM}. As seen, 
the $G'$-boson can give in the forward-backward asymmetry $A_{\rm FB}^{p \bar p}$ the contribution 
of about 10 \% .  

So, the $G'$-boson induced by the chiral color symmetry~\eqref{chiral_group}
in general case of~$g_L\neq g_R$ is consistent with the data~\eqref{expcspptt09},~\eqref{AFBpptt09}
and can reduce the difference between the experimental and SM values~\eqref{AFBpptt09}, ~\eqref{AFBppttSM}
of the forward-backward asymmetry $A_{\rm FB}^{p \bar p}$ in the $t\bar{t}$ production at the Tevatron.

\vspace{5mm}

\centerline{\Large  Summary   }

\begin{itemize} 

\item 
The contributions of $G'$-boson predicted by the chiral color symmetry of quarks
to the cross section $\sigma_{t\bar{t}}$ and to the forward-backward asymmetry
$A_{\rm FB}^{p \bar p}$ of $t\bar{t}$ production at the Tevatron
are calculated and analysed in dependence on two free parameters of the model,
the $G'$ mass $m_{G'}$ and mixing angle $\theta_G$. 

\item 
The $G'$-boson contributions to $\sigma_{t\bar{t}}$ and $A_{\rm FB}^{p \bar p}$ are shown
to be consistent with the Tevatron data
on $\sigma_{t\bar{t}}$ and $A_{\rm FB}^{p \bar p}$ and the allowed region
in the $m_{G'} - \theta_G$ plane is discussed, 
in particular, it is shown that for 
$$m_{G'}>1.02 \, TeV $$ 
in the $m_{G'} - \theta_G$ plane there is the region with~$1 \sigma$ consistency.

\item 
So, the $G'$-boson induced by the chiral color symmetry of quarks   
in general case of~$g_L\neq g_R$ is consistent with the Tevatron data 
on $\sigma_{t\bar{t}}$ and $A_{\rm FB}^{p \bar p}$ 
and can reduce the difference between the experimental and predicted by SM values  
of the forward-backward asymmetry $A_{\rm FB}^{p \bar p}$ in the $t\bar{t}$ 
production at the Tevatron.

\end{itemize}

%
%
The authors are grateful to 
Organizing Commitee of the International Seminar "Quarks-2010"
for possibility to participate in this Seminar.

The work is supported by the Ministry of Education and Science of Russian Federation   
under state contract No.P2496 
of the Federal Programme "Scientific and Pedagogical Personnel of Innovation Russia"  
for 2009-2013 years.



%

\end{document}